\documentclass[pdflatex,sn-apa]{sn-jnl}

\usepackage{graphicx}%
\usepackage{multirow}%
\usepackage{amsmath,amssymb,amsfonts}%
\usepackage{amsthm}%
\usepackage{mathrsfs}%
\usepackage[title]{appendix}%
\usepackage{xcolor}%
\usepackage{subfigure}
\usepackage{textcomp}%
\usepackage{manyfoot}%
\usepackage{booktabs}%
\usepackage{algorithm}%
\usepackage{algorithmicx}%
\usepackage{algpseudocode}%
\usepackage{listings}%

\theoremstyle{thmstyleone}%
\theoremstyle{thmstyletwo}%

\theoremstyle{thmstylethree}%

\begin{document}

\title[The STAR method]{Selecting representative community partitions under modularity degeneracy: the STAR method}


\author[1]{\fnm{Francesca} \sur{Grassetti}}\email{francesca.grassetti@uniurb.it}

\author*[2]{\fnm{Rossana} \sur{Mastrandrea}}\email{rossana.mastrandrea@unito.it}

\affil[1]{\orgdiv{Department of Economics, Society, Politics}, \orgname{University of Urbino}, \orgaddress{\street{via Saffi 42}, \city{Urbino}, \postcode{61029},  \country{Italy}, ORCID: 0000-0001-8703-3232}}

\affil[2]{\orgdiv{Department of Management}, \orgname{University of Turin}, \orgaddress{\street{Corso Unione Sovietica 218bis}, \city{Turin}, \postcode{10134}, \country{Italy}, ORCID: 0000-0001-8596-6389}}


\abstract{Community detection based on modularity maximization is one of the most widely used approaches for uncovering mesoscale structures in complex networks. However, it is well known that the modularity function exhibits a highly degenerate optimization landscape: a large number of structurally distinct partitions attain close modularity values. This degeneracy raises issues of instability, reproducibility, and interpretability of the detected communities.
We propose a simple and user-friendly post-processing method to address this problem by selecting a representative partition among the set of high-modularity solutions. The proposed approach is model-agnostic and can be applied a posteriori to the output of any modularity-based community detection algorithm. Rather than seeking the optimal partition in terms of modularity, our method aims to identify a solution that best represents the structural features shared across degenerate partitions.
We compare our approach with consensus clustering methods, which pursue a similar objective, and show that the resulting partitions are highly consistent, while being obtained through a substantially simpler procedure that does not require additional optimization steps or external software packages. Moreover, unlike standard consensus clustering techniques, the proposed method can be applied to networks with both positive and negative edge weights, making it suitable for a wide range of applications involving signed networks and correlation-based systems, such as social, financial, and neuroscience networks.

Overall, the method provides a practical and robust tool for handling degeneracy in modularity-based community detection, combining simplicity with broad applicability across different types of networks and real-world problems.}




\maketitle

\section{Introduction}
Many real-world systems -- spanning natural, social, technological, and economic domains -- are characterized by complex patterns of interactions among their constituent elements, giving rise to collective behaviors, non-trivial structural organization, and emergent phenomena that cannot be fully understood by analyzing individual components in isolation. By representing such systems as networks, where nodes denote entities and edges encode interactions, graph theory provides a principled approach to model, analyze, and interpret complex interdependencies, enabling insights into system robustness, dynamics, information flow, and functional organization \citep{posfai2016network,newman2018networks}.

 Within this broader context, a fundamental task in network analysis is Community Detection (CD), also known as graph or network clustering \citep{Fortunato2016}. The objective of CD is to partition the nodes of a network into substructures or communities (or clusters/modules), where nodes within a group are densely connected while nodes belonging to different groups are sparsely linked. Identifying these mesoscopic structures is crucial for revealing the latent functions, organizational principles, and operational processes of complex network systems \citep{Jin2023}.
 CD is a versatile network analysis technique applied across numerous scientific and technological domains, with significant impact in fields ranging from social sciences to biology and beyond \citep{Karatas}. Specific application contexts include social networks \citep{Bedi}, political science, neuroscience and genetics \citep{Shai}, 
as well as  economics with the identification of asset clusters in financial markets for portfolio diversification and risk management \citep{almog, bazzi, zanin, musmeci}, the detection of anomalies and fraud in banking and trading networks \citep{safdari, mazzarisi}, the analysis of interlocking networks of firms \citep{heemskerk2013community,drago2017communities,mastrandrea2025context}, the grouping of countries or banks in the interbank lending system and international trade network \citep{barigozzi2011identifying, bargigli, grassi, siudak}.

The problem of CD is inherently ill-defined, lacking a universal and unambiguous definition of what constitutes a \lq\lq community\rq\rq{}. Over the years, numerous methodologies have been proposed. Classical approaches include hierarchical clustering, spectral clustering, and statistical inference \citep{LIneuro}. Statistical modeling is prominent, notably through Probabilistic Graphical Models such as the Stochastic Block Model, which formalizes a generative process based on probability distributions \citep{abbe}. More recently, new research lines have emerged leveraging the capacity of deep learning to handle high-dimensional network data and learn low-dimensional representations of network structures, encompassing methods like those based on auto-encoders and Graph Convolutional Networks \citep{LIUhon, YUAN, Zhang}. Among the optimization-based approaches, \textit{modularity} maximization is the most popular and extensively studied method for CD. Modularity, defined by \citet{Newman2004} and generally indicated with $Q$, is a quality function that measures the fraction of links falling within communities compared to its expected number according to a null mode preserving node connections. 
Finding the global optimum of $Q$ is known to be NP-hard \citep{Brandes2008}. Consequently, heuristic algorithms, such as the Louvain method \citep{demeo} and the Clauset-Newman-Moore algorithm \citep{bulbahce}, are widely employed to find highly modular partitions.

Despite its ubiquity, modularity maximization suffers from critical limitations. Foremost among these is the resolution limit, a bias resulting from the underlying null model that prevents the detection of smaller communities in large networks \citep{Fortunato2007, Lancichinetti2011}. Crucially, the optimization process is highly susceptible to finding degenerate solutions.

Formally, let $G = (V,E)$ be a finite graph and let $\mathcal{P}(V)$ denote the set of all partitions of the vertex set $V$. Modularity maximization consists in solving the discrete optimization problem \[Q_{\max} := \max_{P \in \mathcal{P}(V)} Q(P),\]
where $Q : \mathcal{P}(V) \to \mathbb{R}$ is the modularity functional. This problem is NP-hard and the objective function is neither convex nor submodular over the partition lattice.

Beyond computational intractability, the modularity landscape exhibits a pronounced structural near-degeneracy. In networks composed of $k$ weakly interconnected modular building blocks, the number of structurally distinct partitions attaining objective values arbitrarily close to $Q_{\max}$ grows at least as $2^{k-1}$, and may in general increase super-exponentially with $k$. At the same time, the modularity variation induced by merging or rearranging modules scales as $O(k^{-2})$, implying that large families of partitions differ combinatorially while remaining nearly indistinguishable in objective value \cite{Good_2010}.

To formalize this phenomenon, for $\varepsilon > 0$ let us define the $\varepsilon$-optimal set
\[\mathcal{P}_{\varepsilon}
=
\left\{
P \in \mathcal{P}(V)
\; : \;
Q_{\max} - Q(P) \le \varepsilon
\right\}\]
For sufficiently small $\varepsilon$, the set $\mathcal{P}_{\varepsilon}$ may already have exponential cardinality in the number of mesoscopic structures. Consequently, the global maximizer is typically not isolated but embedded in a high-dimensional plateau of near-optimal solutions.

Importantly, partitions in $\mathcal{P}_{\varepsilon}$ may exhibit substantial structural variability despite negligible differences in modularity. Therefore, modularity maximization does not generically yield a structurally stable solution, and selecting a representative partition within the near-optimal set $\mathcal{P}_{\varepsilon}$ constitutes a secondary optimization and inference problem over a highly degenerate solution manifold.

To address the ambiguity arising from degeneracy and structural variability, several strategies have been proposed. Consensus Clustering (CC) \citep{Lancichinetti2012} aims to merge the information from an ensemble of degenerate partitions  into a single, more stable solution. CC typically uses a consensus matrix based on the co-occurrence of nodes in the same community across the ensemble. However, consensus solutions can sometimes inadvertently blur relevant information captured by individual partitions \citep{cal_19}. An alternative approach is the recursive significance clustering scheme, which identifies subsets of nodes that maintain stable joint community assignments under network perturbation, representing the persistent and well-supported features of the network structure, distinguishable from unstable nodes \citep{Economou2025}. The Bayan algorithm \citep{Aref2023} make use of integer programming to approximate the global maximum of modularity, eliminating degeneracy by ensuring consistent, high-quality partitions - though at higher computational cost. Other methods have also been employed to maximize modularity and detect statistically significant communities and hierarchies, they include the message passing for modularity \citep{Zhang2014}, a proper assignment of weights to the edges \citep{Khadivi}, and the modelling of mixed populations of partitions \citep{Peixoto}.

Despite the substantial body of work addressing degeneracy in modularity-based CD, an open challenge remains the development of a selection criterion that is both simple and broadly applicable, without introducing additional modeling assumptions or significant computational overhead. Moreover, most existing methods implicitly assume non-negative edge weights, thereby excluding signed networks and correlation-based systems. This restriction is particularly limiting in empirical settings where interactions naturally admit both positive and negative values, such as in Social Science and Financial networks.

In this work, we introduce a simple and user-friendly post-processing method aimed at selecting a representative partition from an ensemble of high-modularity solutions. The proposed method is fully agnostic with respect to the underlying CD algorithm and can be applied a posteriori to any modularity-based approach, independently of the specific heuristic or optimization strategy used to generate the candidate partitions. Importantly, the method focuses on extracting a partition that best represents the structural information shared across degenerate solutions.
The rationale behind this approach is that, in the presence of a highly degenerate modularity landscape, the pursuit of an optimal partition is often neither robust nor informative. Small variations in the network or in the optimization process may lead to markedly different solutions with comparable modularity values. In such settings, a representative partition -- capturing the most recurrent and structurally stable features of the ensemble -- provides a more reliable and interpretable description of the mesoscale organization of the network.

We explicitly relate our method to CC \citep{Lancichinetti2012}, as both approaches pursue the same objective of stabilizing community assignments in the presence of degeneracy. However, the proposed framework achieves comparable results through a substantially simpler procedure. Furthermore, our method naturally extends to networks with both positive and negative edge weights, allowing it to be applied to signed networks and correlation-based matrices without any ad hoc modification, a limit of the CC method.

Overall, the contribution of this paper is twofold. First, we provide a parsimonious and easily implementable tool for selecting representative community structures in modularity-based CD, without sacrificing the quality of the resulting partitions. Second, we extend the applicability of representative partition selection to all types of networks, including signed networks, thereby broadening the scope of modularity-based methods across a wide range of theoretical and applied contexts.

The remainder of the paper is organized as follows. Section Method introduces the proposed selection procedure and outlines the associated selection framework. Section Data describes the benchmark and real-world datasets considered in the analysis. Section Results reports the empirical performance of the method on synthetic benchmark networks with known ground truth and on real economic networks. Finally, Sections Discussion and Conclusion summarize the main findings and discuss implications and possible extensions.

\section{Method}

In this section, we describe the methodology adopted to select a representative community partition from an ensemble of modularity-based solutions.

\subsection{Network properties}Let us consider a graph $G=(V,E)$, where $V$ is the set of nodes with size $|V|=N$ and $E$ is the set of edges with size $|E|=L$.
The structure of the graph is encoded in the binary adjacency matrix $A \equiv (a_{ij})_{1\le i,j\le N}$, where $a_{ij} = 1$ if an edge exists from node $i$ to node $j$, $0$ otherwise. For directed networks, in general $a_{ij} \neq a_{ji}$. When edges are associated with weights, the network is characterized by the weighted adjacency matrix $W \equiv (w_{ij})_{1\le i,j \le N}$, where $w_{ij}$  denotes the weight of the edge from node $i$ to node $j$. As in the binary case, weights may be asymmetric in directed networks \citep{posfai2016network,newman2018networks}.

In a binary directed network, each node $i$ is characterized by its out-degree and in-degree, defined respectively as
\[
k^{\mathrm{out}}_i = \sum_{j} a_{ij}, \qquad 
k^{\mathrm{in}}_i = \sum_{j} a_{ji},
\]
which measure the number of edges leaving and entering node $i$, respectively.

In a weighted directed network, the corresponding quantities are the out-strength and in-strength, defined by
\[
s^{\mathrm{out}}_i = \sum_{j} w_{ij}, \qquad
s^{\mathrm{in}}_i = \sum_{j} w_{ji},
\]
representing the total weight exiting from and incoming to node $i$, respectively.
The total number of edges in the network - $L$ - and the total volume of exchanges - $w_{\mathrm{tot}}$ - can be represented in terms of degree and strength:
\[
L = \sum_{i,j} a_{ij} = \sum_i k^{\mathrm{in}}_i = \sum_i k^{\mathrm{out}}_i,
\quad \quad \quad 
w_{\mathrm{tot}} = \sum_{i,j} w_{ij} 
= \sum_i s^{\mathrm{out}}_i
= \sum_i s^{\mathrm{in}}_i ,
\]
\subsection{Modularity maximization} Modularity is one of the most widely used quality functions for CD, providing a quantitative criterion for evaluating how well a given partition captures the mesoscale organization of a network. The modularity–maximization approach \citep{Newman2004,Newman2006} does not require specifying the number of communities \textit{a priori}. Instead, it attempts to automatically uncover the mesoscale structure of the network based on two fundamental assumptions: (i) nodes belonging to the same community are more likely to be connected than nodes in different communities, and (ii) a random network constructed under appropriate constraints does not exhibit any intrinsic community structure. Modularity thus measures the extent to which the observed density of intra–community connections exceeds what would be expected under a suitable null model.
Formally, modularity is defined as
\begin{equation}
Q=\frac{1}{L}\sum_{i,j\in V}\left(a_{ij}-p_{ij}\right)\delta(c_i,c_j),
\label{mod}
\end{equation}

where the Kronecker delta $\delta(c_i,c_j)$ equals $1$ if nodes $i$ and $j$ belong to the same community and $0$ otherwise. The term $p_{ij}$ introduces the notion of \textit{randomness}, representing the expected probability of an edge between nodes $i$ and $j$ under a benchmark null model preserving selected structural properties of the observed network.
In binary and weighted networks, the most commonly adopted null models are, respectively, the following: 

\begin{equation}
p_{ij} = \frac{k^{\mathrm{in}}_i , k^{\mathrm{out}}_j}{L}, \quad 
p_{ij} = \frac{s^{\mathrm{in}}_i , s^{\mathrm{out}}_j}{w_{\mathrm{tot}}}
\label{nullmodel}
\end{equation}

corresponding, respectively, to the expected
number of edges joining nodes $i$ and $j$ by rewiring links  to preserve the degree and the expected edge-weight between nodes $i$ and $j$ preserving the strength of all vertices, on average.

Modularity maximization formulates CD as a global optimization problem in which one searches, over all possible partitions of a network, for the division of vertices that yields the highest value of a modularity function (\ref{nullmodel}). Modularity is in practice maximized using approximate methods. Widely used approaches include greedy and multilevel agglomerative schemes that iteratively merge nodes or communities to increase modularity \citep{Newmangreedy}, spectral algorithms based on the eigenvectors of the modularity matrix \citep{Richardson}, 
metaheuristic strategies such as extremal optimization \citep{Brandes2008} or simulated annealing \citep{Guimerà_2005} that stochastically explore the space of partitions, and mathematical programming formulations that provide exact or approximate solutions \citep{Chen}.

For our purpose, we adopted the Louvain algorithm \citep{Blondel2008} to address the computational complexity of modularity maximization. The Louvain method is a greedy, hierarchical optimization procedure that is specifically designed to handle large networks efficiently. It iteratively improves modularity through local node movements and successive aggregation of communities, enabling the detection of community structure at multiple scales. Therefore, it offers a favorable balance between accuracy, scalability, and computational efficiency. Finally, it is applicable to both weighted and unweighted networks making it particularly well suited for the network structures analyzed in this study.\\

\subsection{Similarity-based Top ARI Representative method}

We introduce the Similarity-based Top ARI Representative (STAR) method to address the degeneracy inherent in modularity maximization.
The method is fully agnostic with respect to the specific optimization heuristic used to generate candidate partitions and can be applied to the output of any modularity-maximization algorithm. 
For concreteness, in the following we generate candidate partitions by repeatedly applying the Louvain algorithm.

Given the graph $G$, we repeat $T$ times the modularity maximization procedure using the Louvain algorithm  obtaining $T$ optimal partitions $\boldsymbol{\sigma} = \{\sigma_1, \dots, \sigma_T\}$ and  corresponding modularities $\mathbf{Q} = \{Q(\sigma_1), \dots, Q(\sigma_T)\}$.

We employ the Adjusted Rand Index (ARI) \citep{hubert1985comparing} to quantify pairwise similarities between the $T$ partitions of the graph $G$. ARI  evaluates the agreement between two partitions by considering all pairs of nodes and correcting for chance agreement, yielding an expected value of zero for random and independent partitions. This property is particularly important when comparing multiple solutions generated by stochastic and degenerate optimization procedures such as modularity maximization.

Alternative measures commonly used in CD include the Normalized Mutual Information (NMI)  \citep{danon2005comparing} and the Variation of Information (VI)  \citep{meilua2007comparing}. Although NMI is widely adopted, it does not explicitly correct for chance agreement and is known to be biased by the number and size of communities, potentially leading to inflated similarity values when comparing unrelated partitions \citep{vinh2009information}. VI, on the other hand, defines a metric distance between partitions , but lacks a natural baseline for random agreement and is less intuitive when interpreted as a similarity measure. In contrast, ARI is invariant under permutation of community labels, accommodates partitions with differing numbers of clusters, and provides a robust and interpretable notion of similarity, making it well suited for identifying a representative solution among degenerate partitions.

Let $\sigma_i$ and $\sigma_j$ be two partitions with cluster sets $\mathcal{C}^{[i]} = \{C^{[i]}_1, \dots, C^{[i]}_l\}$ and $\mathcal{C}^{[j]} = \{C^{[j]}_1, \dots, C^{[j]}_y\}$. For each pair of clusters $(C^{[i]}_u, C^{[j]}_v)$, define $n_{uv} = \bigl|C^{[i]}_u \cap C^{[j]}_v\bigr|$, representing the number of nodes assigned simultaneously to cluster $u$ in partition $i$ and cluster $v$ in partition $j$. Summing across rows and columns gives the marginal totals:
\[
n_{u\cdot} = \sum_{v=1}^{y} n_{uv}, \qquad
n_{\cdot v} = \sum_{u=1}^{l} n_{uv}.
\]

Using these quantities, let us define
\[
z = \sum_{u=1}^{l} \sum_{v=1}^{y} \binom{n_{uv}}{2}, \quad
b = \sum_{u=1}^{l} \binom{n_{u\cdot}}{2}, \quad
c = \sum_{v=1}^{y} \binom{n_{\cdot v}}{2}, \quad
M = \binom{N}{2}.
\]

The ARI is then computed as
\[
\operatorname{ARI}(\sigma_i,\sigma_j) = \frac{z - \frac{bc}{M}}{\frac{1}{2}(b+c) - \frac{bc}{M}}
\]
In this formula, \(z\) counts the number of pairs of nodes that are clustered together in both partitions, while \(b\) and \(c\) count the total number of pairs in each partition separately. The term \(bc/M\) represents the expected number of agreements between the partitions under a random assignment of nodes to clusters. The ARI corrects for this expected chance agreement, yielding a value of 1 when the partitions are identical, 0 when the similarity is no better than random, and negative values when the partitions agree less than expected by chance. This makes the ARI particularly suitable for comparing partitions of different sizes or cluster counts, and it treats cluster splitting and merging symmetrically.

Pairwise ARI values are used to build a weighted undirected network in which nodes represent the $T$ partitions of the graph $G$ and edge weights correspond to the pairwise ARI between them and whose adjacency matrix is indicated by $R\equiv (r_{ij})_{1 \le i,j \le T}$.
We can compute the strength of each node in this network as \[
s_i = \sum_{i=1}^T r_{ij} = \sum_{j=1}^T r_{ij}
\]
Since the similarity network is fully connected by construction, the strength of the node can be interpreted as an average measure of how similar a given partition is to the remaining $T-1$ partitions\footnote{Indeed, it is sufficient to divide the strength of each node by
$T-1$, after setting the main diagonal of the ARI matrix--corresponding to self-similarity values -- to zero.}.

Within this framework, the algorithm first identifies the partitions with maximal strength and then selects, among these, the partition(s) with the highest modularity value. This procedure is designed to identify the partition(s) that are most \emph{representative} of the entire ensemble while simultaneously favoring solutions that are closest to the optimal modularity, thereby balancing representativeness and quality.

We stress that the goal of the STAR method is not to improve the optimum of the objective function. Global quality functions such as modularity are known to exhibit strong degeneracy \citep{Fortunato2007,Lancichinetti2011}, and insisting on the absolute optimum often does not lead to meaningful or stable partitions \citep{Lancichinetti2012}. STAR instead selects a representative solution among degenerate partitions, focusing on consistency rather than optimization through a simple and user-friendly approach. 

For clarity, the main steps of the proposed procedure are summarized in the pseudo-algorithm reported below.

\begin{algorithm}[H]
\caption{\textbf{Pseudo-algorithm:} Similarity-based Top ARI Representative (STAR) method}
\begin{algorithmic}[1]
\Require Graph $G$, number of repetitions $T$

\Ensure Representative partition of $G$ with high modularity

\State Repeat modularity maximization $T$ times on $G$
\State Collect resulting partitions $\boldsymbol{\sigma} = \{\sigma_1, \dots, \sigma_T\}$ and modularities $\mathbf{Q} = \{Q(\sigma_1), \dots, Q(\sigma_T)\}$

\State Compute pairwise similarities between partitions using ARI
\State Construct a fully connected weighted network of partitions (nodes correspond to the $T$ partitions, edges to the pairwise ARI)

\State Compute the strength of each partition as the sum of its edge weights
\State Identify the partition(s) with maximal strength
\State Among these, select the partition(s) with the highest modularity value

\end{algorithmic}
\end{algorithm}

In the following, we assess the performance of the proposed method on both synthetic benchmark networks with known ground truth and on real-world networks.

\section{Data}

In this section, we describe the datasets used to evaluate the proposed method. 
We consider two complementary types of data: synthetic benchmark networks with known ground-truth community structure, and real-world economic networks.
This combination allows us to assess the performance of the method under controlled conditions and to illustrate its behavior in applied settings.

\subsection{Benchmarks with ground-truth}

We evaluate the proposed approach on artificial benchmark graphs with planted community structure using the LFR benchmark \citep{Lancichinetti2008}, a standard framework for assessing clustering algorithms. LFR graphs reproduce key properties of real networks, including power-law distributions of node degree and community size. A central quantity is the mixing parameter $\mu \in [0,1]$, which controls the fraction of links each node shares outside its own community. Lower values of $\mu$ correspond to well-defined communities, while higher values indicate increasing overlap between communities, sharply reducing the performance of modularity-based methods. We consider weighted, undirected networks of size $N=1000$ and generate 100 graph instances for each value of $\mu \in \{0.1, 0.2, \ldots, 0.9\}$.

This benchmark setting allows us to directly evaluate the quality of the selected representative partition by comparing it with the known ground-truth community structure.

\subsection{Real Data}
We apply our procedure to two economic datasets: international trade relationships among world countries and equity market data from the FTSE 100 index.

International trade data are obtained from the CEPII database\footnote{https://cepii.fr/CEPII/en/bdd\_modele/bdd\_modele\_item.asp?id=8}
. Each observation reports the exporter, the importer, the deflated trade flow (in billions of US dollars), and the reference year. We focus on the year 2015, which includes the largest number of countries (193) within the 2000–2020 period. The resulting World Trade Web (WTW) is modeled as a network whose nodes represent countries and links are weighted with the total value of exported and imported goods and services.

The second dataset consists of daily equity price time series for firms included in the FTSE 100 index. Data were retrieved from Refinitiv via the LSEG Workspace platform under an academic license from University of Turin. For each stock, daily adjusted closing prices were collected, accounting for corporate actions such as stock splits and dividend payments. We retain 2,544 trading days per stock, covering approximately ten years from April 1, 2016 to December 31, 2025. Stocks with excessive missing observations were excluded, while remaining gaps were filled using forward-fill imputation. Each firm was assigned to a sector based on the Datastream Industrial Sector Classification (Level 2), mapped to the Industry Classification Benchmark. Log-returns were computed and pairwise correlations were estimated, yielding a correlation matrix that was subsequently filtered using Random Matrix Theory and eigendecomposition techniques to remove market-wide and random effects \citep{macmahon2015community}. 

These datasets were selected for two main reasons. First, they represent structurally different networks: the WTW is weighted and directed, whereas the FTSE 100 network is based on a correlation matrix (symmetric and semidefinite positive) that includes negative values, making it unsuitable for the CC method. Second, both datasets illustrate applied contexts in which CD serves as a preliminary step for further analysis, underscoring the importance of a simple and reliable partition selection method.

\section{Results}
In this section we present the empirical performance of the STAR method. We first validate the procedure on synthetic benchmark networks with planted community structure (LFR graphs), assessing both accuracy with respect to the ground truth and the associated modularity values. We then illustrate the method on two real-world economic networks -- the WTW and the FTSE100 correlation network -- to highlight its behavior in applied settings, including cases with signed weights where standard consensus approaches are not directly applicable.

\subsection{Benchmark}
The purpose of the benchmark analysis is to evaluate the ability of the STAR method to select a representative partition that is both structurally meaningful and consistent with the underlying community organization, in a setting where the ground truth is known. Synthetic benchmarks allow us to isolate the effect of partition selection from confounding factors and to systematically investigate how performance changes as communities become increasingly mixed.\\
To this end, we rely on LFR benchmark networks, which are specifically designed to reproduce key statistical properties of real-world networks -- such as heterogeneous degree distributions and community sizes -- while providing a well-defined planted partition. For each value of the mixing parameter $\mu \in \{0.1, 0.2, \ldots, 0.9\}$, which controls the fraction of inter-community links and thus the level of community fuzziness, we generate 100 independent networks with 1000 nodes. All remaining parameters are fixed across realizations in order to isolate the effect of increasing $\mu$. The chosen parameter configuration follows standard practice in the literature and is consistent with previous benchmark studies on CC \citep{Lancichinetti2012}, ensuring that observed differences are not driven by idiosyncratic settings.\\
For each network realization, we generate an ensemble of candidate partitions by running the Louvain algorithm 150 times. This step reflects the intrinsic stochasticity and degeneracy of modularity maximization: even on the same network, repeated optimizations typically yield multiple near-optimal but structurally distinct solutions.
From each ensemble of partitions, a single representative solution is extracted using the STAR method and, for comparison, the CC.

\begin{figure}[!ht]
    \subfigure[]
    {\includegraphics[width=0.5\textwidth]{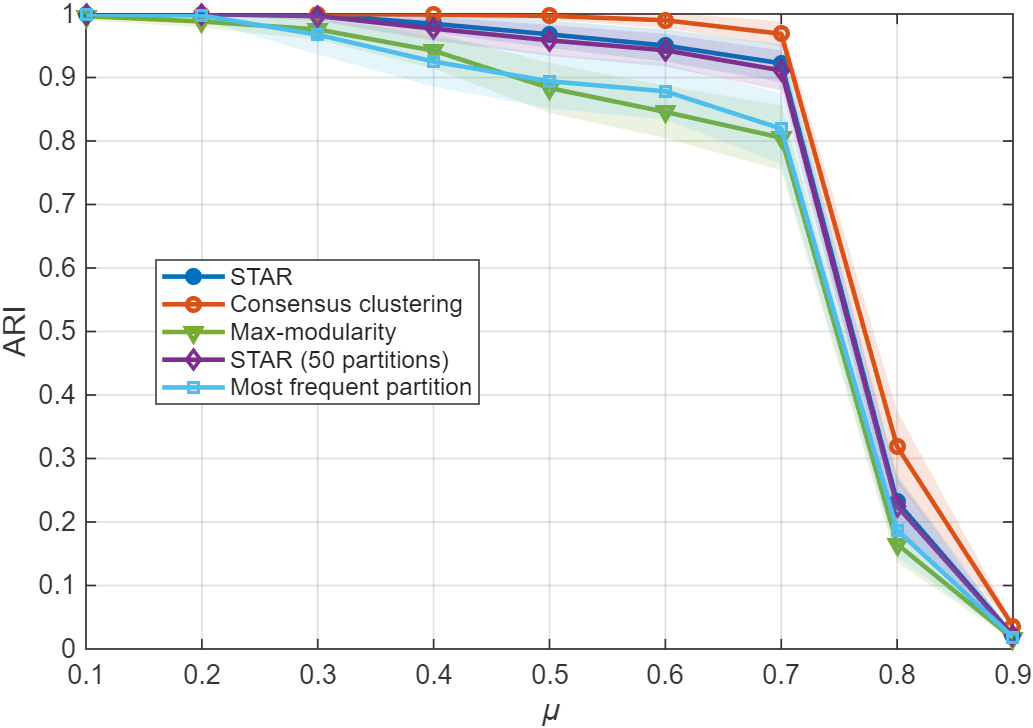}}
    \subfigure[]
     {\includegraphics[width=0.5\textwidth]{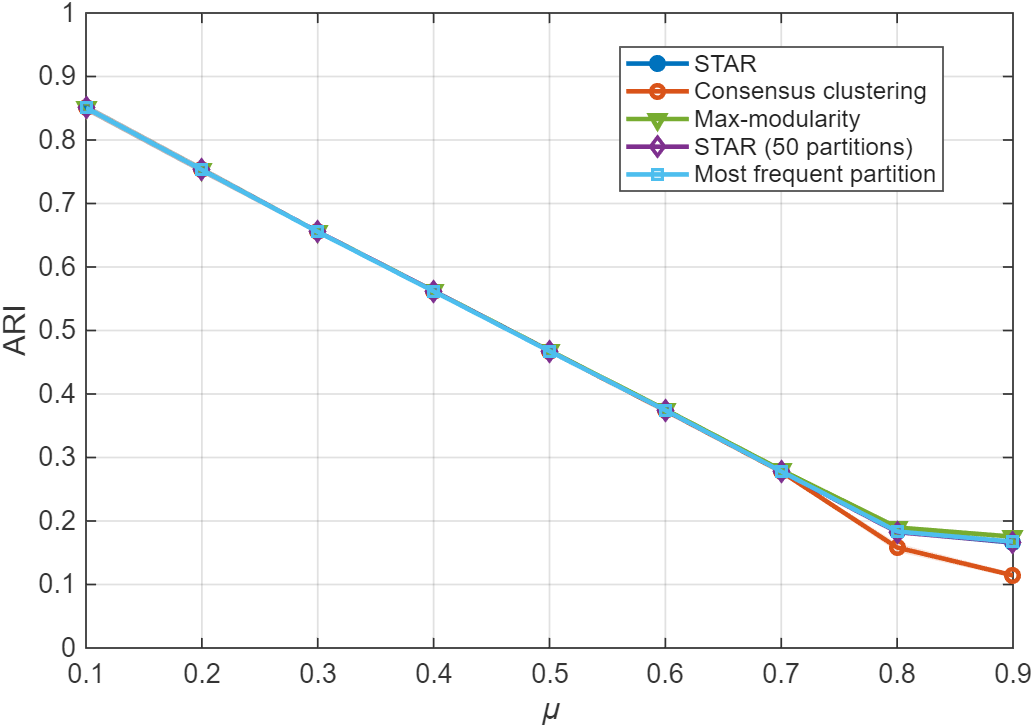}}
    \caption{The parameters of the LFR benchmark graphs are: average degree = 20, maximum degree = 50, minimum community size = 10, maximum community size = 50, degree exponent = 2, and community size exponent = 3.}
    \label{fig:1}
\end{figure}

Once a representative partition is selected for each network realization, its quality is assessed by comparing it with the planted ground-truth partition using ARI. For each value of $\mu$, this results in 100 ARI values per method. Following standard practice in the community-detection literature \citep{fortunato2010community}, we summarize these results by reporting the mean ARI  as a function of $\mu$. This allows us to explicitly track how the accuracy of the representative partition deteriorates as communities become less well defined. The corresponding curves are reported in Fig. \ref{fig:1}, panel (a) together with the standard deviation (shaded areas). In parallel, we evaluate the modularity associated with the representative partitions selected by each method. For each value of $\mu$, mean modularity values are computed over the 100 realizations and reported in Fig. \ref{fig:1}, panel (b). This second set of results is crucial to assess whether selecting a representative partition comes at the cost of a substantial loss in modularity, or whether high representativeness can be achieved while remaining close to the modularity optimum.

To further contextualize the benchmark results, we additionally consider teo purely modularity-driven selection criteria, whereby the representative partition is chosen as the one with the highest modularity and the one appearing most frequently among the 150 Louvain runs. For these baselines, we compute both the ARI with respect to the ground truth and the corresponding modularity values, and include the resulting averages and standard deviations in Fig. \ref{fig:1}, panels (a)–(b). These comparisons highlight the extent to which maximizing modularity alone or choosing the most frequent partition as the most representative one may lead to clusters that are less faithful to the underlying community structure, despite achieving marginally higher modularity scores.

Finally, we investigate the robustness of the STAR method with respect to the size of the partition ensemble. Specifically, we repeat the entire procedure using only 50 Louvain runs instead of 150. 

The results provide a clear picture of how different selection criteria behave as the level of community mixing increases. As shown in panel (a), for low values of  $\mu$, all methods achieve similarly high ARI values, indicating that when communities are well separated the degeneracy of modularity maximization is limited and different representative partitions largely coincide.
As $\mu$ increases and the planted community structure becomes progressively less pronounced, differences between selection strategies emerge. In this regime, CC achieves the highest agreement with the ground truth. The STAR method closely tracks the consensus performance across the entire range of $\mu$, with only marginal deviations, showing that STAR is able to capture essentially the same structural information despite relying on a much simpler selection mechanism. Notably, reducing the number of input partitions from 150 to 50 has a negligible impact on STAR performance, as both ARI values and standard deviations (shaded areas) remain virtually unchanged. This indicates that the method is robust and does not require large ensembles of candidate partitions to yield reliable results.
In contrast, selecting the partition solely based on maximum modularity, as well as choosing the most frequently occurring partition among Louvain runs, leads to systematically poorer agreement with the planted structure.

Panel (b) reports the modularity values associated with the representative partitions. For low and intermediate values of $\mu$ (up to approximately $\mu = 0.7$), all methods yield almost identical modularity scores. For larger values of $\mu$,  CC exhibits slightly lower modularity values, although differences occur where modularity is already low for all methods.

Overall, the benchmark analysis shows that the STAR method provides a reliable and robust criterion for selecting representative partitions in modularity-based CD. It achieves accuracy comparable to CC while maintaining modularity values close to the optimum, and remains effective even with a substantially reduced number of input partitions. These results support the use of STAR as a practical alternative for representative partition selection in settings characterized by strong modularity degeneracy.

\subsection{Real Data}

After validating the proposed selection procedure on LFR benchmark networks, we apply the method to two real-world systems, namely the WTW and the FTSE100 correlation network.In the WTW case, communities are identified using the Louvain algorithm, producing an ensemble of 150 partitions. We then select a representative solution according to three criteria: CC, STAR, and the partition attaining the maximum modularity. In the second case, the presence of negative weights prevents the application of CC; hence, only the STAR partition and the maximum-modularity partition are considered.

\subsubsection{The World Trade Web}
The WTW considered in this application refers to the year 2015 and includes 193 nodes, representing countries, connected by 26,015 directed and weighted links. The resulting network is highly dense, with a link density of approximately 0.7, and encodes a total trade volume of $1.19\times10^{10}$
 US dollars. These structural features are known to pose challenges for CD, particularly for modularity-based approaches.
Indeed, multiple candidate partitions are obtained from repeated Louvain runs. The STAR selection criterion and the consensus-based approach consistently identify the same partition (hereafter STAR/Consensus partition), while the partition maximizing the standard modularity function (Max-mod partition) differs from the STAR-consensus solution. This discrepancy highlights the well-known tendency of modularity maximization to favor large and weakly structured communities in dense and highly heterogeneous networks such as International Trade.

\begin{figure}[!ht]
    \subfigure[STAR/Consensus partition ($Q=0.2649$)]    {\includegraphics[width=0.9\textwidth]{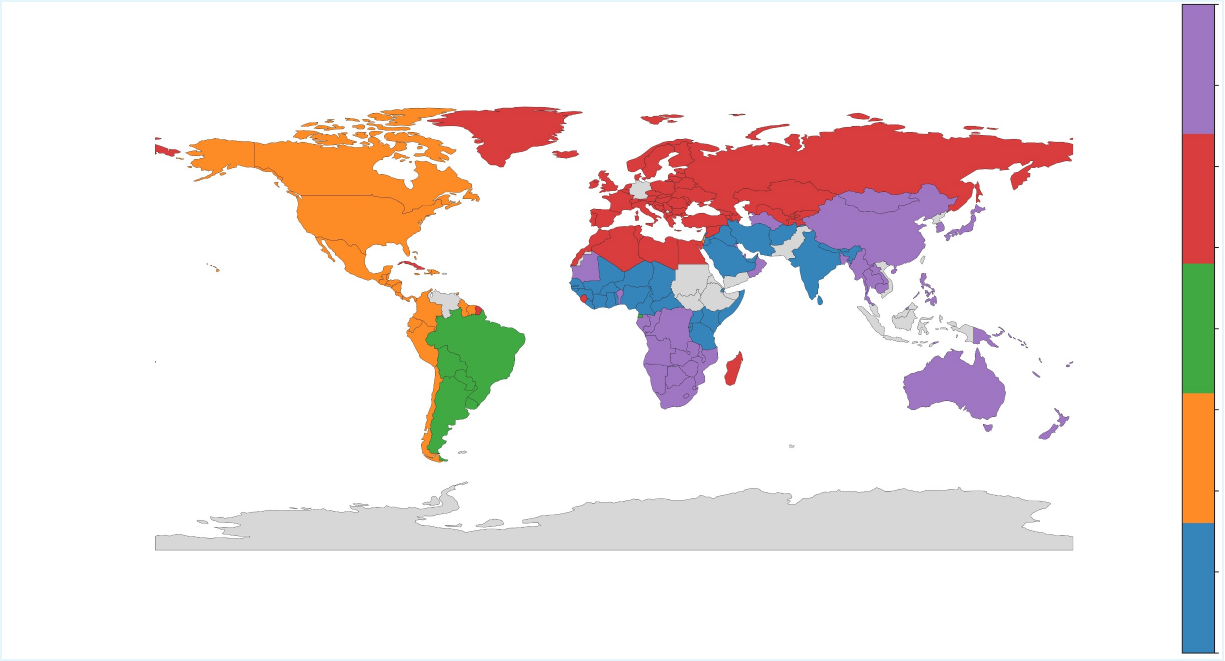}}
    \subfigure[Max-mod partition ($Q=0.2720$)]     {\includegraphics[width=0.905\textwidth]{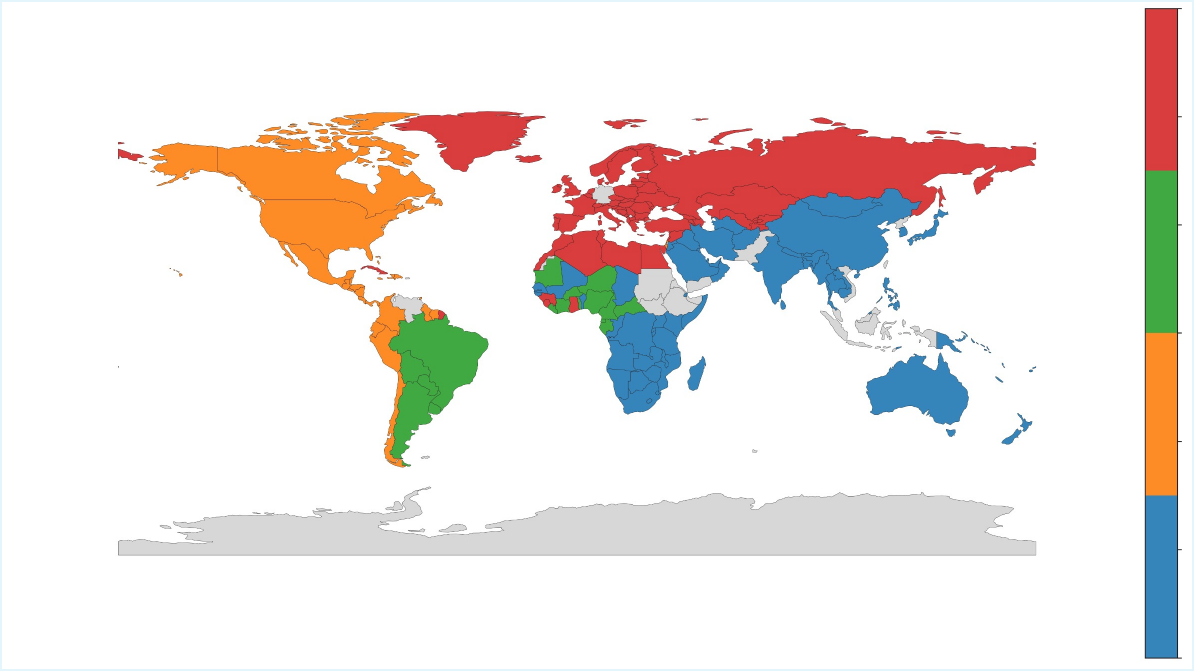}}
    \caption{World Trade Web 2015. Countries are colored according to the communities identified by (a) the STAR/consensus method and (b) the maximum modularity criterion.}
    \label{fig:WTW2015}
\end{figure}

The STAR/Consensus partition displays a geographically coherent and economically interpretable community structure. North America forms a unified block, South America is largely grouped into a single cluster centered on Brazil and its regional trade partners, and Europe appears as a well-defined and compact community. Asia is split into two major blocks, broadly distinguishing East Asia and Oceania from South and Central Asia, while Sub-Saharan Africa exhibits a differentiated structure consistent with known trade patterns. Overall, the partition reflects well-documented regional trade integration patterns in the WTW, such as the strong intra-European trade core, the North American integration, and the regional clustering of South American economies. In contrast, the maximum-modularity partition appears less aligned with established economic geography. While Europe and North America remain coherent clusters, large parts of Africa and Asia are aggregated into broader and less economically homogeneous communities. In particular, several African economies are merged into a single large block despite substantial heterogeneity in their trade linkages, and parts of Asia are grouped in a way that blurs the distinction between major regional trade hubs. Although this partition achieves the highest modularity value, it does so at the cost of reduced economic interpretability and weaker correspondence with known regional trade agreements and trade intensity patterns.

Taken together, these results illustrate the well-known degeneracy of the modularity landscape: partitions with very similar modularity values may differ substantially in their structural composition. In the WTW case, the STAR-selected partition appears more consistent with established trade blocs and regional integration patterns, suggesting that maximizing modularity alone does not necessarily yield the most economically meaningful community structure.

\subsubsection{FTSE100 stock market} 
The second application concerns the FTSE100 equity market, represented as a weighted network of stocks, where nodes correspond to companies included in the FTSE100 index and links encode pairwise correlations between their return time series. Such financial networks are known to exhibit strong heterogeneity and high levels of interconnectedness, making them a relevant test case for CD and selection methods .We selected all stocks for which complete data are available during the period under study ending up with $93$ stocks. They can be classified in ten top-level "sectors": Basic Materials,
    Consumer Discretion, 
    Consumer Staples,
    Energy,
    Financials,
    Health Care,
    Industrials
    Technology,
    Telecommunications,
    Utilities.

\begin{figure}[!ht]
    \centering
   \subfigure[Histogram of modularity values.] {\includegraphics[width=0.6\textwidth]{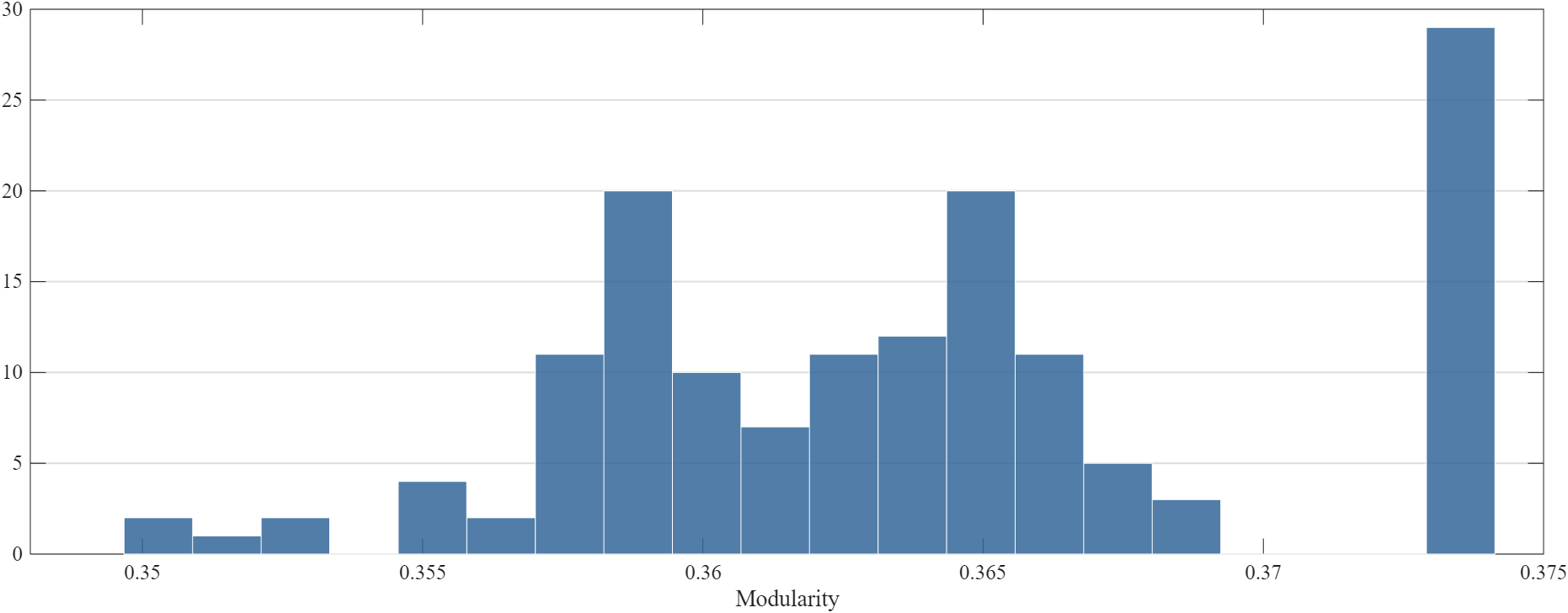}}  
   \subfigure[STAR partition ($Q=0.3654$).] {\includegraphics[width=0.5\textwidth]{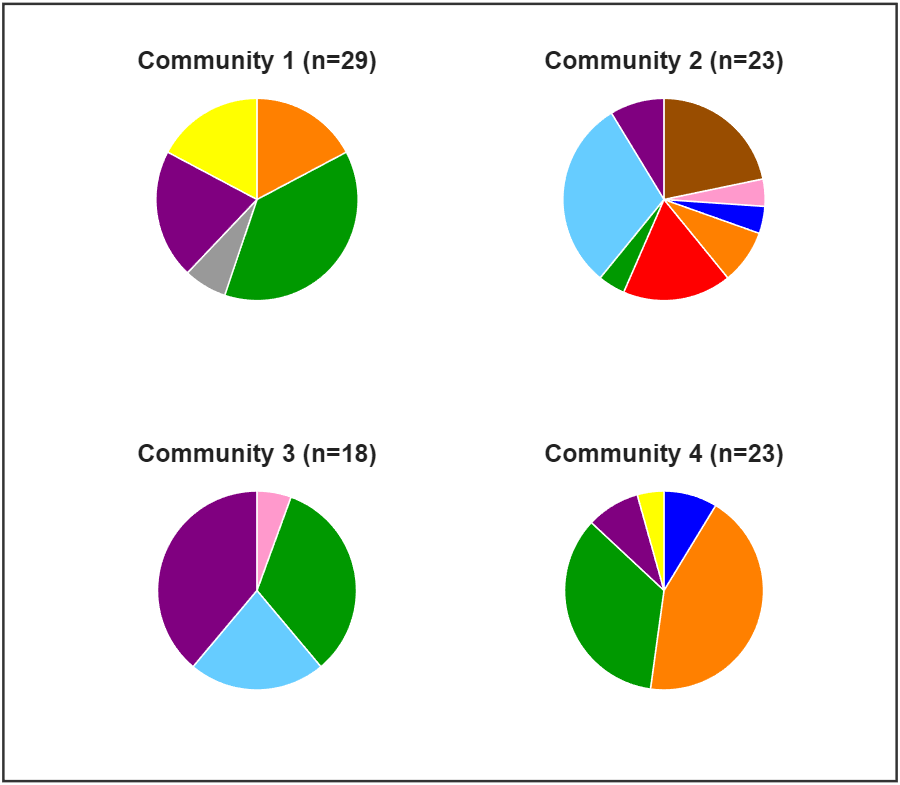}}   
\subfigure[Maximum-modularity partition ($Q=0.3741$).]{\includegraphics[width=0.6\textwidth]{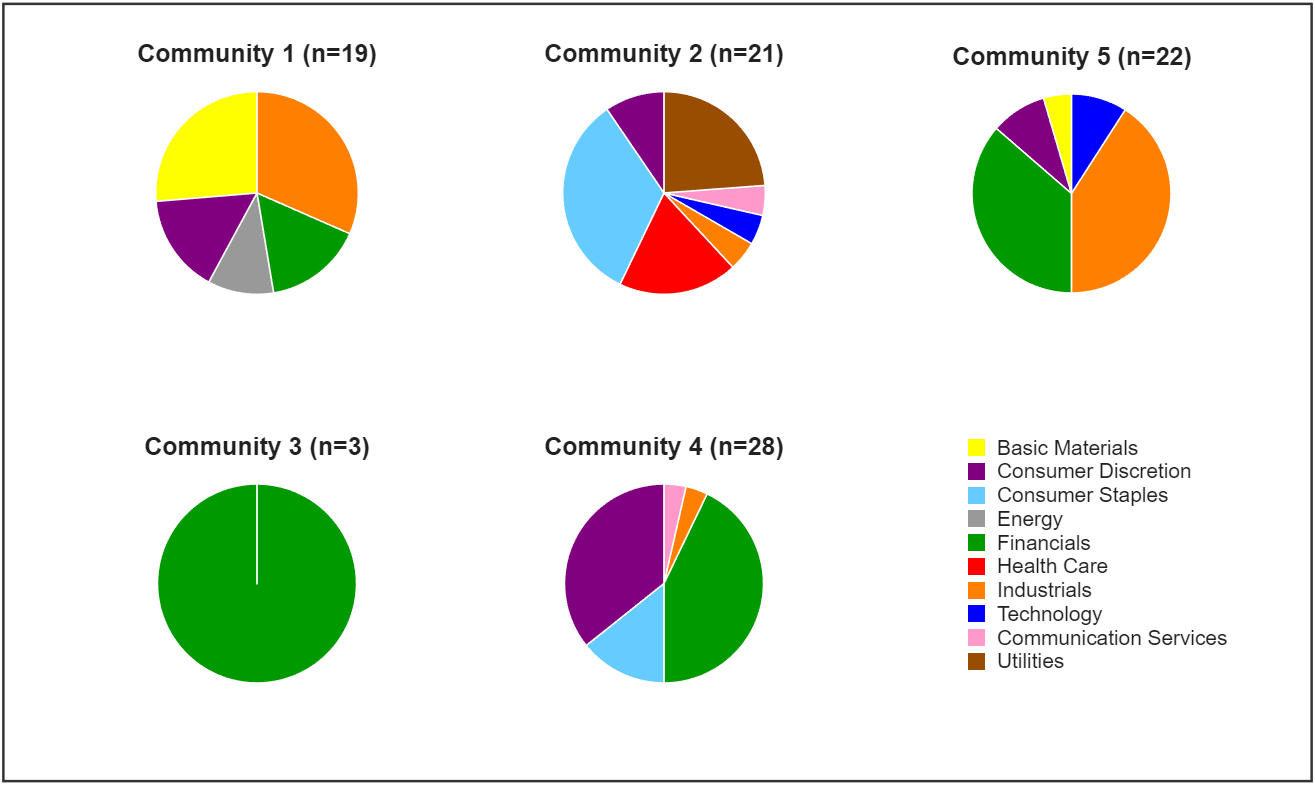}}
    \caption{\textbf{Communities of FTSE100 market.}  Partitions are generated using the Louvain algorithm on the filtered correlation matrix (REF) for 150 runs. (a) Histogram of the modularity values associated to the 150 partitions. Community organization in sectors of (b) the representative partition obtained with the STAR method and (c) the partition associated to the maximum modularity value.}
    \label{fig:FTSE100}
\end{figure}

The filtered correlation matrix contains negative values, which makes standard CC procedures not directly applicable and motivates the use of STAR in this setting. In Figure \ref{fig:FTSE100}, panel (a), we report the histogram of modularity values over 150 partitions obtained with Louvain algorithm confirming the degeneracy problem. The STAR method returns a representative partition that differs from the one associated with the maximum modularity value among the 150 detected partitions. The difference in modularity is negligible ($\Delta Q = 0.0087$); however, the resulting community structures differ both in number and composition, with STAR identifying four communities and the maximum-modularity solution yielding five clusters (Figure \ref{fig:FTSE100}, panels (b)–(c)).

The STAR partition exhibits a clearer sectoral organization (Figure \ref{fig:FTSE100}, panel (b)). Financial firms are more coherently grouped, Industrials form a dominant and well-defined block, and defensive sectors such as Consumer Staples and Utilities tend to cluster together. Overall, the resulting communities display lower internal sectoral heterogeneity and align more closely with standard GICS classifications and with well-established patterns of sectoral co-movement in equity markets. This suggests that the STAR procedure selects a partition that is structurally central within the ensemble and economically interpretable in terms of common risk exposures.

By contrast, the maximum-modularity partition, while optimizing the graph-theoretic objective function, produces a more fragmented sectoral structure (Figure \ref{fig:FTSE100}, panel (c)). Although Financials remain prominent, several communities mix heterogeneous sectors, and sectoral concentration appears less pronounced. This is consistent with the fact that modularity maximization identifies an extremal solution in the objective landscape, which does not necessarily coincide with the most representative or economically coherent configuration when the modularity surface is highly degenerate.

Taken together, the comparison highlights that, in correlation-based financial networks, maximizing modularity and selecting a structurally central partition may lead to different outcomes. In such near-degenerate settings, the STAR criterion provides a partition that better reflects known economic structure, whereas the maximum-modularity solution strictly optimizes the objective function without imposing additional stability considerations, while CC is not appliable.

\section{Discussion}
In this paper we introduce a simple and user-friendly post-processing method designed to support the selection of a representative partition among the degenerate solutions produced by modularity-based CD algorithms. The proposed approach is deliberately model-agnostic: it can be applied a posteriori to any CD method relying on modularity maximization and is independent of the specific optimization heuristic employed. Moreover, it is applicable to any network topology, including weighted, directed, and signed networks.

The motivation for developing such a method stems from well-known limitations of modularity maximization. Despite its widespread use, the modularity function is characterized by a highly rugged optimization landscape, where an exponential number of structurally distinct partitions may exhibit similar or nearly identical modularity values. As a consequence, the notion of a single “best” partition -- identified solely as the global or local maximizer of modularity -- is often ill-defined. In practice, optimal or near-optimal solutions may be unstable across runs, sensitive to algorithmic randomness, or even associated with community structures that are difficult to interpret from a modeling or empirical perspective.

For this reason, our approach does not aim at identifying the best partition in terms of modularity value. Instead, it seeks a representative partition, namely a solution that captures the most robust and recurrent structural features of the ensemble of high-modularity partitions. This conceptual shift reflects the idea that, in complex networks, a meaningful description of the system should prioritize structural consistency and interpretability over marginal improvements in an objective function. Similar considerations have motivated the development of CC methods \citep{Lancichinetti2012}, which aim to extract a stable partition summarizing a set of competing solutions.
We explicitly compare our method with CC because both approaches pursue the same overarching goal: selecting a partition that is representative of the underlying network structure rather than arbitrarily choosing among degenerate optima. Our results show that the partitions obtained with the proposed method are highly consistent with those produced by CC, both in terms of community composition and structural coherence. Importantly, this is achieved through a substantially simpler procedure, which does not require the construction of auxiliary networks or the use of additional optimization steps, and can be implemented using standard outputs of modularity-based algorithms without relying on external packages.

Beyond simplicity, a key advantage of the proposed method lies in its generality. Standard CC techniques are typically defined for non-negative edge weights and therefore cannot be directly applied to networks with negative weights. This limitation excludes a wide class of relevant applications, including signed social networks, correlation-based networks in finance, neuroscience, and climatology, as well as any setting in which positive and negative interactions coexist. In contrast, our method naturally extends to networks with both positive and negative weights, thereby providing a unified framework for partition selection across a broad spectrum of real-world problems.

Overall, the contribution of this work is twofold. First, we offer a parsimonious and easily implementable tool for selecting representative community structures without sacrificing the quality of the results typically obtained by more elaborate consensus-based approaches. Second, we extend the applicability of representative partition selection to all types of networks, including signed networks, thus broadening the scope of modularity-based CD in both theoretical and applied contexts. 

\section{Conclusion}
In this paper we addressed one of the central challenges of modularity-based community detection, namely the degeneracy of high-modularity partitions and the resulting ambiguity in the selection of a meaningful community structure. Rather than pursuing the identification of a single optimal solution, we proposed a simple post-processing method aimed at selecting a representative partition that captures the most robust structural features shared across degenerate modularity-maximizing solutions.
The proposed approach can be applied a posteriori to the output of any modularity-based community detection method. By focusing on representativeness rather than optimality, the method provides a stable and interpretable alternative to the selection of a “best” partition in contexts where the modularity landscape is highly rugged and near-optimal solutions are structurally diverse.
A key strength of the proposed framework lies in its simplicity and generality. Compared with consensus clustering, which pursues a similar objective, our method achieves comparable results through a substantially simpler procedure. Moreover, unlike standard consensus-based approaches, the method naturally extends to networks with both positive and negative edge weights, making it suitable for signed networks and correlation-based systems that arise in a wide range of applications, including social, financial, and neuroscience networks.\\
The numerical experiments confirm that the representative partitions selected by the proposed method are structurally coherent, robust across realizations, and consistent with those obtained through more elaborate stabilization techniques. These results suggest that the method provides a practical and reliable tool for handling degeneracy in modularity-based community detection without sacrificing the quality of the inferred community structure.

\backmatter

\bmhead{Acknowledgements} R.M acknowledges support from the PRIN 2022 project “The Role of the
Public and Private Sectors in Pharmaceutical Breakthrough Innovations (3PBI)” (CUP
2022S4EAS9). R.M. is member of GNAMPA (Gruppo Nazionale per
l’Analisi Matematica, la Probabilit`a e le loro Applicazioni) at INdAM (Istituto Nazionale
di Alta Matematica).

\section*{Declarations}

\begin{itemize}
\item \textbf{Funding.} No funding was received for conducting this study.
\medskip
\item \textbf{Conflict of interest/Competing interests.}The authors have no relevant financial or non-financial interests to disclose.
\medskip

\item \textbf{Data availability.} Trade data are publicly available from the CEPII database \url{https://cepii.fr/CEPII/en/bdd_modele/bdd_modele_item.asp?id=8}. \\Financial data are accessed through a personalized license agreement with the LSEG database at the University of Turin and cannot be shared due to licensing restrictions.
\medskip

\item \textbf{Code availability.} \url{https://github.com/RossanaMastrandrea/STAR-partition}
\medskip

\item \textbf{Author contribution.} The authors contributed equally to this work.
\end{itemize}







\bibliography{sample}

\end{document}